\documentclass[preprint,showpacs,preprintnumbers,amsmath,amssymb]{revtex4}

\usepackage{graphicx}% Include figure files
\usepackage{dcolumn}% Align table columns on decimal point
\usepackage{bm}% bold math
\usepackage[]{ams,amssymb,epsfig,color}
\newcommand{\be}{\begin{equation}}
\newcommand{\ee}{\end{equation}}
\begin{document}
%\preprint{JPhysB}
\title{Dynamic switching of magnetization in driven magnetic molecules}
%Manuscript Title:\\with Forced Linebreak}% Force line breaks with \\
%\title{Dynamics of dipolar molecular chain in crossed static: Soliton formation}
%Manuscript Title:\\with Forced Linebreak}% Force line breaks with \\

\author{L. Chotorlishvili$^{1,3}$, P. Schwab$^{1}$, J. Berakdar$^{2}$}
%
%\institute{
%  \inst{1} Institut f\"ur Physik, Martin-Luther Universit\"at Halle-Wittenberg, Heinrich-Damerow-Str.4
%06120 Halle, Germany\\
%\inst{2}Institute f\"ur Physik, Universit\"at Augsburg, 86135 Augsburg, Germany\\
% \inst{3} Physics Department of the Tbilisi State University,
%                 Chavchavadze av.3, 0128, Tbilisi, \\ Georgia }
%  \inst{2} Second Institute - Address

\affiliation{1 Institut f\"ur Physik, Universit\"at Augsburg, 86135 Augsburg, Germany\\
   2 Institut f\"ur Physik, Martin-Luther Universit\"at
Halle-Wittenberg, Heinrich-Damerow-Str.4, 06120 Halle, Germany \\
  3 Physics Department of the Tbilisi State University,
                 Chavchavadze av.3, 0128, Tbilisi, Georgia}

\begin{abstract}
We study the magnetization dynamics    of  a molecular
 magnet  driven by static and variable magnetic fields within a semiclassical treatment. The underling analyzes is
 valid in a regime, when \ the energy
 is  definitely lower than  the anisotropy barrier, but still a substantial number of states
are excited. We find   the phase space  to contain a separatrix
line. Solutions far from it are  oscillatory whereas
  the separatrix solution is of a soliton type. States near the separatrix
 are extremely sensitive to  small perturbations, a fact which we utilize for dynamically
induced magnetization switching.
\end{abstract}

%Valid PACS numbers may be entered using the \verb+\pacs{#1}+
%command.
\maketitle

%\section{\label{Sec:Introduction} Introduction}

\section{Introduction}
Molecular  magnets (MM) are molecular structures with a large
effective spin ($S$), e.g. for the prototypical MM $Mn_{12}$
acetates \cite{1j} $S=10$.   MM show a number of interesting
phenomena that have been in the focus of  theoretical
   and experimental research \cite{1,1j,2j,3j,4j,2,3,4,5,6,7}.
   To name but few,  as a result of the strong uniaxial
anisotropy,   MM show a bistable behavior
    \cite{1j}; they also exhibit a resonant
tunnelling of magnetization \cite{1}  that shows up as steps in
the magnetic hysteresis loops \cite{2j,3j,4j}. Of special
relevance for applications in quantum computing is  the  large
relaxation time of MM \cite{8}.

The present theoretical work focuses on the dynamics of the
  magnetization. The established picture of
macroscopic quantum tunnelling of the magnetization
 is as follows: The MM effective spin Hamiltonian $\hat{H}=-DS^{2}_{z}$
 possesses  degenerate
energy levels $\pm M_{S},-S<M_{s}<S$ separated by the finite
barrier $E_{B}=DS^{2}$.
%In spite of this, experimental evidence suggests that
%the spin system is able to overcome this barrier by quantum
%tunnelling \cite{1}.
At low temperatures only the lowest levels $M_{S}=\pm S$ are
populated. Those two states are orthogonal to each other and no
tunnelling is possible. An anisotropic perturbation
$E(S_{x}^{2}-S_{y}^{2})$ does not commute with the Hamiltonian
$\hat{H}=-DS^{2}_{z}$  and  mixes therefore the states at  both
sides of the anisotropy barrier leading thus to tunnelling
\cite{4}. Reversal of the magnetization due to macroscopic quantum
tunnelling has a maximum for the states close to the top of the
barrier.  This case corresponds  to the high temperature limit.

 In this work we consider the
magnetization dynamics  induced by constant and harmonic external
magnetic fields: The influence of a variable magnetic field on MM
at
  low temperatures, i.e. when  only 2-3 levels are
excited was considered in \cite{6,7}. It was shown that in this
case the problem is reduced  to a three level Jaynes-Cummings
model, the so called Lambda configuration. Therefore, it is
analytically solvable in principle. The low temperature assumption
is however quite restrictive \cite{8, Hill}: If only the levels
$E^{0}$, $E^{1}$, $E^{2}$ are involved the low temperature
approximation  is applicable for temperatures obeying
 \cite{7}
$k_{B}T<E^{1}-E^{0}$, where $k_B$ is the Boltzmann constant. For
$Mn_{12}$ this leads to the  estimate  $T<0.6K$ \cite{Hill}.
Obviously, if the temperature exceeds $T$, an approximation with a
large number of levels  participating in the process is more
appropriate.
% (For illustration, spin of
%$Mn_{12}$, $S=10$, $S=16$
%).
 In this case the
quasi-classical approximation for the spin dynamics becomes
applicable \cite{9,10}. It is our aim here to conduct such a
study. MM will be modelled as in previous  studies, e.g. in  Refs.
\cite{6,7}. We consider the dynamical reversal of the
magnetization, caused not by an anisotropic perturbation  but by a
constant and varying magnetic field. The energy is such that a
large number of levels are excited,
 but still low enough such that tunnelling induced by an
 anisotropic perturbation is weak.
%Due to
%nonlinearity of the system dynamics promise to be nontrivial.
%Investigation of this problem is an aim of this paper. In
%particular,
We shall show that under different conditions (depending on the
fields parameters), different types of magnetization dynamics are
realized. For the time evolution of the magnetization vector,
under certain conditions we obtain  a solution of the soliton
type. The various types of the dynamics will be linked to the
structure of the phase space of the system. In particular, the
 existence of
the separatrix in the phase space has a profound influence on the
system behavior. We will show that in this case a new type of
field-assisted
  magnetization dynamics emerges, namely a dynamically induced switching. This occurs when  the energy of the
system (in the presence of the field) is still lower than the
re-scaled anisotropy barrier \cite{11} and coincides with the
separatrix values of the energy. Therefore, the domain close to
the separatrix is identified as the phase space area where the
dynamically induced switching takes place.
\section{Model}
We  consider a molecular magnet, e.g. $Fe_{8}$ or
$Mn_{12}$ acetate. The uniaxial anisotropy axis (easy axis)
 sets the  $z$- direction. The MM is   subjected to a constant magnetic field
directed along  the  $x$-axis and a radio frequency (rf) magnetic
field polarized in the $x-y$-plain. The Hamiltonian of the single
molecular magnet reads \cite{7}
\begin{eqnarray}
&&\hat{H}=\hat{H_{0}}+\hat{H_{I}} ,\nonumber\\
&&\hat{H_{0}}=-D\hat{S_{z}^{2}}+g\mu_{B}H_{0}\hat{S_{x}},\\
&&\hat{H_{I}}=-\frac{1}{2}g\mu_{B}H_{1}e^{i\omega_{0}t}(\hat{S_{y}}+\hat{S_{x}})+H.c. \, .\nonumber
\end{eqnarray}
Here $D$ is the longitudinal anisotropy constant, $\hat{S_{x}}$,
$\hat{S_{y}}$, $\hat{S_{z}}$ are the projections of the spin operators
along the  $x,y,z$ axis,
$g$ is the Land\'e factor,  and $\mu_{B}$ is the Bohr magneton.
 $H_{0}$
stands for the constant magnetic field amplitude whereas  $H_{1}$,
and $\omega_{0}$ are the amplitude and the frequency of the rf
field. The problem when both fields $H_{0},~H_{1}$ are time
dependent was studied in \cite{12}. Using quantum-mechanical
perturbation theory, the probability of quantum tunnelling of
magnetization has been estimated. However, here we are interested
in the exact solution of the semi-classical equations of motion.
     Typical values of the parameter $D$  are $90$  GHz for $Mn_{12}$ and
$D=30$ GHz for $Fe_{8}$ \cite{Hill, 10}.
 Since we are interested in the case when large number of levels are exited, the spin of the magnetic
 molecule can be treated as a classical vector on the Bloch sphere.
 Taking into account that  $S^{2}=S_{x}^{2}+S_{y}^{2}+S_{z}^{2}$
 is an  integral of motion, it is appropriate to switch to
the
 new variables $(S_{z},~\varphi)$ via the transformation [14]: $S_{x}=\sqrt{1-S_{z}^{2}}\cos\varphi$, $S_{y}=\sqrt{1-S_{z}^{2}}\sin\varphi$ and rewrite
  (1) in the compact form: \be
H=-\frac{\lambda}{2}S_{z}^{2}+\sqrt{1-S_{z}^{2}}
\cos\varphi-\varepsilon\sqrt{1-S_{z}^{2}}(\sin\varphi+\cos\varphi)\cos(\omega_{0}t)
. \ee Hereafter, if not otherwise stated the energy  and the time
scales are set by the constant magnetic field $H \mapsto
H/g\mu_{B}H_{0}S,~~t \mapsto \frac{2DS}{\lambda}t,~~\omega_{0}
\mapsto \frac{\lambda}{2DS} \omega_{0}$. We introduced two
dimensionless parameters $\lambda=\frac{2DS}{g\mu_{B}H_{0}}$,
$\varepsilon=\frac{H_{1}}{H_{0}}<1$. The corresponding Hamilton
equations are
\begin{eqnarray}
&& \dot{S_{z}}=-\frac{\partial H}{\partial
\varphi}=\sqrt{1-S_{z}^{2}}\sin\varphi+\varepsilon\sqrt{1-S_{z}^{2}}(\cos\varphi-\sin\varphi)\cos(\omega_{0}t),\nonumber
\\
&&\dot{\varphi}=\frac{\partial H}{\partial S_{z}}=-(\lambda+
\frac{\cos\varphi}{\sqrt{1-S_{z}^{2}}})S_{z}+\varepsilon\frac{S_{z}}{\sqrt{1-S_{z}^{2}}}(\sin\varphi+\cos\varphi)\cos(\omega_{0}t).
\end{eqnarray} These equations are nonlinear. Therefore,  the
solutions to (3) can be regular or chaotic, depending on the
values of the magnetic fields (parameters $\lambda$,
 $\varepsilon$). From the intuitive point of view it is obvious,
that for the low energy case, i.e. close to the ground states
$S_{z}\approx\pm 1$, the system Eq.(2) should become linear.
However in the language of variables action angle
$(S_{z},~\varphi)$ that is not so trivial. Therefore, we will
discuss this question in more details when studying solutions of
the autonomous system.

\section{Autonomous system: An exact solution}
We inspect at first the autonomous system, i.e. when $\varepsilon=0$.
 In this case the system can be integrated exactly:
Taking into energy conservation  $H=const=-\Sigma$ \be
\frac{\lambda}{2}S_{z}^{2}-\sqrt{1-S_{z}^{2}}\cos\varphi=\Sigma
\ee and \be \dot{S_{z}}=\sqrt{1-S_{z}^{2}}\sin\varphi,\ee we find
 \be
 \dot{S_{z}^{2}}+\big[\frac{\lambda
 S_{z}^{2}}{2}-\Sigma\big]^{2}=1-S_{z}^{2} .
\ee
    Consequently from eq.(6) we infer
    \be
    \frac{\lambda
    t}{2}=\int\limits_{S_{z}(t)}^{S_{z}(0)}\frac{dS_{z}}{\sqrt{\big(\frac{2}{\lambda}\big)^{2}\big(1-S_{z}^{2}\big)-\big[S_{z}^{2}-\frac{2\Sigma}{\lambda}\big]^{2}}}.
    \ee
    This relation can be rewritten in the
 form \be    \frac{\lambda
t}{2}=\int\limits_{S_{z}(t)}^{S_{z}(0)}\frac{dS_{z}}{\sqrt{\big(a^{2}+S_{z}^{2}\big)\big(b^{2}-S_{z}^{2}\big)}},
\ee where
$a^{2}=\frac{2}{\lambda^{2}}\big[\theta^{2}/2-(\Sigma\lambda-1)\big]$,
$b^{2}=\frac{2}{\lambda^{2}}\big[\theta^{2}/2+(\Sigma\lambda-1)\big]$,
$\theta^{2}(\lambda)=2\sqrt{\lambda^{2}-2\Sigma\lambda+1}$.
Performing the integration (8) and inverting the result
  we obtain
\begin{eqnarray}
S_{z}(t)=\left\{\begin{array}{ll} b\;
 \textrm{cn}[(b\lambda/k)(t-\alpha),~~k],~~~~~~
0<k<1, \\
 b\;
\textrm{dn}[(b\lambda/k)(t-\alpha),~~1/k],~~~ k>1.
\end{array}
 \right.
\end{eqnarray}
Here $\textrm{cn}(...)$ and $\textrm{dn}(...)$ are the Jacobi
periodic functions. The coefficients that enter  eq. (9)  read
\begin{eqnarray}
&&k^{2}=\frac{1}{2}\bigg(\frac{b\lambda}{\theta(\lambda)}\bigg)^{2}=\frac{1}{2}\bigg[1+\frac{(\Sigma\lambda-1)}{\sqrt{\lambda^{2}+1-2\Sigma\lambda}}\bigg],\nonumber
\\
&&\alpha=2\big[\lambda\sqrt{a^2+b^2}F(\arccos[S_{z}(0)/b],k)\big]^{-1}.
\end{eqnarray}
With
$F(\varphi,k)=\int\limits_{0}^{\varphi}dq(1-k^2\sin^21)^{-1/2}$
being the incomplete elliptical integral of the first kind. From
eq.(9) we conclude that, depending on the values of the parameter
$k$ (10), the dynamics of the magnetization is described by
different solutions. They are separated by the special value $k=1$
of the bifurcation   parameter $k$  indicating thus the presence
of topologically distinct solutions.
%The existence of two solutions separated by a bifurcation
%value of the parameter is very important. We shall come back to this
%question later, but at first we analyze briefly obtained results
%(9).
    In eq.(9) the Jacobian elliptic functions $\textrm{cn}(\varphi,k)$ and
$\textrm{dn}(\varphi,k)$ are periodic in the argument $\varphi$
with the period $4K(k)$ and $2K(k)$ respectively, where
$K(k)=F(\pi/2,k)$ is the complete elliptic integral of the first
kind \cite{13}. The time period of the oscillation of the
magnetization $S_{z}(t)$ is given by
\begin{eqnarray}
T=\left\{\begin{array}{ll} \frac{4kK(k)}{b\lambda}~~~~~~~~\textrm{for} ~~~~~~
0<k<1, \\ \frac{2kK(1/k)}{b\lambda}~~~~~~ \textrm{for}~~~~~~k>1.
\end{array}
 \right.
\end{eqnarray}
If $k\longrightarrow 1$, the period becomes infinite because
$K(k)\longrightarrow\ln(4/\sqrt{1-k^2})$. The evolution in this
special case is given by the non-oscillatory soliton solution
\be
S_{z}(t)=b/\cosh[b\lambda(t-\alpha)] .\ee
Considering eq. (10), we infer that  the bifurcation value of the parameter
$k=1$ is connected with an initial energy of the system via the ratio
 \be
\Sigma_{S}=-H_{S}/g\mu_{B}H_{0}=1,~~~
 H\big(S_{z}(t=0);\varphi(t=0)\big)=-
g\mu_{B}H_{0}=H_{S}.\ee
If this condition (13) is not fulfilled the dynamics of the
magnetization is described by the solutions (9). Finally to
conclude this section we consider linear limit of solutions
eq.(9):

$$\textrm{cn}(u,k)\approx \textrm{cos}(u)+k^{2}\textrm{sin}(u)(u-\frac{1}{2}\textrm{sin}(2u)),~~~k^{2}\ll 1,$$
and
$$\textrm{dn}(u,k)\approx 1-\textrm{sin}(u)^{2}/k^{2},~~~k^{2}\gg1 .~~~~~~~~~~~~~~~~~~~~~~$$
The interpretation of those asymptotic solutions is clear. First
one corresponds to the case when in the effective magnetic field
$H_{eff}=(g\mu_{B}H_{0},0,-DS_{z})$, the $x$- component is
dominant. Therefore the magnetization vector performs small
oscillations $|S_{z}(t)|<1$ trying to be aligned along effective
magnetic field. While in the second case, corresponding to the
ground state solution (system is near to the bottom of double
potential well) the effective magnetic field is directed along the
$z$- axis.
   \section{Topological properties of solutions }
   As established \cite{14,15},  the existence of a  bifurcation parameter indicates
that the solutions separated by it, have different topological
properties. Therefore, it is instructive  to consider the
properties of the solutions (9) in the phase plane.
 The existence of the integral of motion (4) in the autonomous
 case makes it possible to express $S_{z}$ as a function of
 $\varphi$:$~~ S_{z}(\varphi,\Sigma)$. The phase portrait of the system is
 shown in Fig.(\ref{Fig:1}):
 \begin{figure}[t]
 \centering
  \includegraphics[width=12cm]{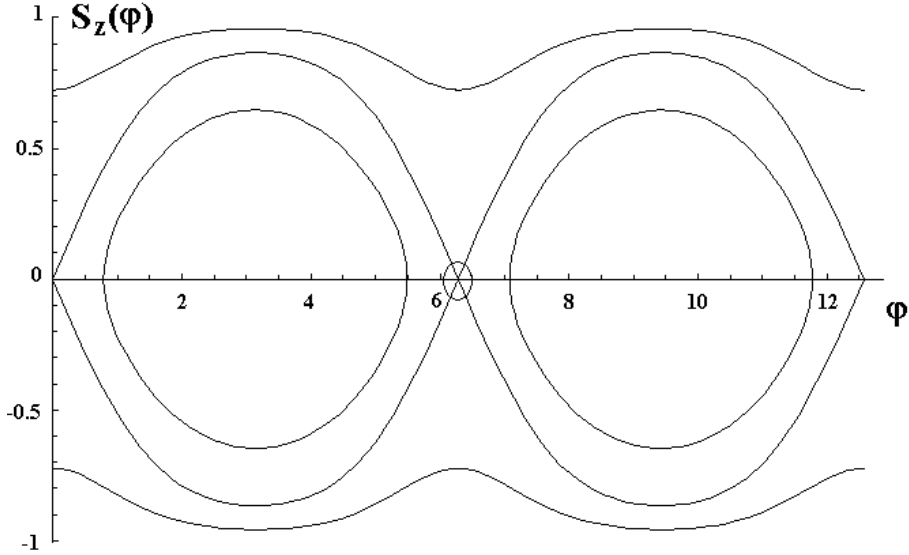}
  \caption{Two types of phase trajectories of the system separated by the separatrix $k=1,~~\Sigma=\Sigma_{S}$. The open
  trajectory (solution $S_{z}(t)=\textrm{dn}(t,1/k)$, $k=1.52,~~\Sigma>\Sigma_{S}$ ) corresponds to the rotational regime of motion.
  The closed trajectory (solution $S_{z}(t)=\textrm{cn}(t,k)$, $k=0.89,~~~\Sigma<\Sigma_{S}$ ) to the oscillatory regime.
 The separatrix crossing point $\bigotimes$ is of special interest:  around this point any perturbation leads to the formation of homoclinic structure.} \label{Fig:1}
\end{figure}
The different phase trajectories correspond to the solutions (9).
The phase trajectories corresponding to the solution
$S_{z}(t)=\textrm{dn}(\varphi,k)$, $k>1$ are open  and they
describe a rotational motion of the magnetization. Trajectories
corresponding to $S_{z}(t)=\textrm{cn}(\varphi,k)$, $k<1$ are
closed and they describe the oscillatory motion of the
magnetization. Closed and open phase trajectories are separated
from each other by the special line called separatrix. The
existence of a separatrix is insofar important as the states in
the phase-space area near the separatrix are very sensitive
\cite{15} to external perturbations, which signals the onset  of
 chaotic behavior.
     The role of perturbations  in our particular case is played
by the applied periodic magnetic field.
 We recall  that the stochastic layer has finite size and it occupies
a  small part of phase space.
 %The determination of the width of the stochastic layer is very important.
% Since motion inside this layer in not regular and it hinders
% controlling of orientation of magnetization.
 \section{ Formation of a stochastic layer}
\begin{figure}[t]
 \centering
  \includegraphics[width=8cm]{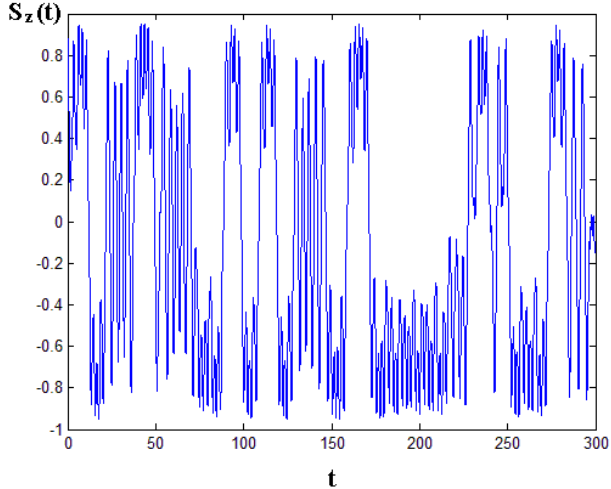}
  \caption{Chaotic motion near the separatrix ($k=1$, $\Sigma=\Sigma_{S}=1$), $D=90GHz$.
  Time independent field $H_{0}$, is chosen such that $\lambda=\frac{2DS}{g\mu_{B}H_{0}}=4$, and
  the ratio between the time independent and variable fields is $\varepsilon=H_{1}/H_{0}=0.3$.
  The initial energy $H=-4.5\cdot10^{3}GHz$ is 8/9 of the re-scaled barrier height $E_{B}^{'}=DS^{2}\big(1-\frac{1}{\lambda}\big)^{2}$.
  Frequency of the variable field is $\omega_{0}=5$. One observes that the
  orientation of the magnetization is changing in time chaotically.} \label{Fig:2}
\end{figure}

 To determine the width of the stochastic layer we follow Ref.\cite{15}.
 For details of the formation of the stochastic layer and for the general
 formalism we refer to the monograph \cite{15}. Here we
 only present the main findings.
We introduce the canonical variable of action
$I=\frac{1}{\pi}\oint S_{z}(\Sigma,\varphi)d\varphi$ and rewrite
the driven nonlinear system (2) in the following form: \be
H=H_{0}+\varepsilon V(I,\varphi)\textrm{cos}(\omega_{0} t). \ee
Here
$H_{0}=\omega(I)I,~~\omega(I)=\bigg[\frac{dI(\Sigma)}{d\Sigma}\bigg]^{-1}$.
The trajectories laying far from separatrix of the unperturbed
Hamiltonian $H_{0}$
 are not influenced by perturbation.
 %Other question
%is trajectories near to the separatrix.
%Key point is that
The motion near the homo-clinic points of the separatrix  is very
slow \cite{15}. Because the period of motion described by (11) is
logarithmically divergent, even small perturbations end up with a
finite influence due to the large period of motion. Thus, the
equations of motion for the canonical variables $(I, \varphi)$

\be \dot{I}=\frac{\partial I}{\partial
H_{0}}\dot{H}=-\frac{\varepsilon}{\omega(I)}\frac{\partial
V}{\partial S_{z}}\dot{S_{z}}\textrm{cos}(\omega_{0} t),\ee

\be \dot{\varphi}=\frac{\partial H}{\partial
I}=\omega(I)+\varepsilon\frac{\partial V}{\partial
S_{z}}\dot{S_{z}} \textrm{cos}(\omega_{0} t), \ee may be
integrated  taking into account the features of the motion near to
the separatrix. Namely, the  acceleration $\dot{S_{z}}$ gives a
nonzero contribution in the integral $ \int dt \frac{\partial
V}{\partial I}\dot{S_{z}}cos(\omega_{0} t)$ only near to the
homoclinic points \cite{15} (the particle moves along the phase
trajectory very fast and spends most of the time near the
homoclinic points). Therefore, the differential equations
(15),(16) can be reduced to the following recurrence relations:
\be
\overline{I}=I-\frac{\varepsilon}{\omega(I)}\int\limits_{\Delta
t}dt\frac{\partial V}{\partial
S_{z}}\dot{S_{z}}\textrm{cos}(\omega_{0}t), \ee \be
\overline{\varphi}= \varphi+\frac{\pi
\omega_{0}}{\omega(\overline{I})}.~~~~~~~~~~~~~~~~~~~~~~~~~~~\ee
Here $\overline{I},\overline{\varphi},$ and $I,\varphi$ are the
values of the canonical variables just after and before passing
the homoclinic point, $\Delta t$ is the interval of the time where
$\dot{S_{z}}$ is different from zero.
 One can deduce the
coefficient of stochasticity by evaluating the maximal Lyapunov
exponent for the Jacobian matrix \be \left( \begin{array}{c}
\frac{\partial\overline{I}}{\partial I}~~\frac{\partial\overline{I}}{\partial \varphi} \\
\frac{\partial\overline{\varphi}}{\partial I}~~\frac{\partial\overline{\varphi}}{\partial \varphi}\\
\end{array} \right),\ee
of the recurrence relations (17),(18). All of this subsume  to the
following expression for the width of the stochastic layer
 \be
  K_{0}=\frac{\pi\varepsilon\omega_{0}}{\omega^{2}}\big|\frac{d\omega}{dH}\big|.
  \ee
Here $\varepsilon$, $\omega_{0}$ are the amplitude and the
frequency of the perturbation. Note that the expression (20) is
general \cite{15} and the only thing one has to do is to calculate
the nonlinear frequency $\omega(I)$ and its derivative with
respect to the energy for the particular system.
 Thus, even  for small perturbation (in our case
  it is the magnetic field with  the frequency $\omega_{0}$ and the amplitude $\varepsilon$, see Eq.(2)) the
  dynamics near the separatrix $k=1$, $H_{c}=-g\mu_{b}H_{0}$ is chaotic
  and unpredictable. Consequently, the solutions (9)  have no meaning
  near the separatrix. At the same time far from the separatrix
  $H \neq H_{S},~~~\Sigma\neq 1$, $k\neq1$ they are valid.
 We note that the expression (20) is
 valid for a low frequency perturbation $\omega_{0} \ll D$ and for a
 high frequency perturbation  $\omega_{0} \geqslant D$ as well. For estimation of the width of the
 stochastic layer $K_{0}$ the variable of action should be determined.
 Taking into account (4) we find
 \be
 I^{\pm}(\Sigma)
= \oint\bigg[\frac{1}{2\lambda^{2}}\bigg(2\lambda
\Sigma-\cos^{2}\varphi\pm2\lambda\cos\varphi
\sqrt{1+\frac{1}{4\lambda^{2}}\cos^{2}\varphi}\bigg)\bigg]^{1/2}d\varphi
.\ee If  the static magnetic field is weak then
$\lambda=\frac{2DS}{g\mu_{B}H_{0}} \gg 1$ is a large parameter.
 In this limit, we can
 simplify expression (21) to
  obtain
 \be
 I(\Sigma)=I^{+}(\Sigma>1)=I^{-}(\Sigma>1)=2\sqrt{\frac{\Sigma+1}{\lambda}}E\bigg(\frac{2}{\Sigma+1}\bigg),
 \ee
where $E(k)$ is  the complete elliptic integral of the second
kind. Taking into account (22) the expression for the width of the
stochastic layer acquires the following form \be
K_{0}\approx\frac{\pi\varepsilon\omega_{0}}{\sqrt{\lambda(\Sigma+1)}|\Sigma-1|}
                 K\bigg(\frac{2}{\Sigma+1}\bigg)E\bigg(\frac{2}{\Sigma+1}\bigg).\ee
 Condition $K_{0}>1$ of the emergence of stochasticity imposes  certain
 restrictions on the parameters of the magnetic field $\varepsilon$,
 $\omega$, $H_{0}$ and on the initial energy $\Sigma$ of the system. When
 the energy approaches the separatrix value $\Sigma\longrightarrow 1$ the
 condition $K_{0}>1$ becomes valid even for a very small $\varepsilon \ll 1$ perturbation. This
 testifies the fact that the system  near the separatrix is sensitive
  to small perturbations. The emergence  of chaos is proved by
 numerical  calculations as well, see  Fig.(\ref{Fig:2}). As one can
 see from this plot, the dynamics is not regular. The projection
 $S_{z}(t)$ of magnetization changes orientation in a chaotic manner.

However, a chaotic change of orientation is not  a reversal to a
stationary target state.
 Under dynamical switching we understand here the transition between the
oscillatory and the rotational types of motion. To be more
specific let us discuss the geometrical aspects of the motion for
the trajectories near the separatrix. Upon applying a static
magnetic field, the magnetization precessional motion in our case
is markedly different from that  in the standard NMR set up:
 The key issue is that the
effective magnetic field $H_{eff}=(g\mu_{B}H_{0},0,-DS_{z})$, due
to the nonlinearity of the system, depends on the values of
$S_{z}$. The magnetization vector tends to align as dictated by
the effective field. However, the orientation of effective field
changes in as much as $S_{z}$ does.  Only in the special case
$S_{z}=0,\varphi=0,2\pi$ which corresponds  to the homoclinic
points the magnetization vector tends parallel to the effective
field $\overrightarrow{M}||\overrightarrow{H_{eff}}$. On the other
hand, the homoclinic point is an unstable equilibrium point.
Therefore, the influence of the variable field leads to a
switching between the two types of the solutions (9). Hence the
following  scenario emerges:
 Suppose at the initial time
the  system is prepared in  the degenerated ground state $M_{s}=
S$. We  apply a constat magnetic along  the $x-$ axis and tune its
amplitude  to realize the separatrix condition
$\Sigma_{s}=g\mu_{b}H_{0}$. A small perturbation can then lead to
the transitions. In particular, switching off the perturbation we
end up with the transformed state (cf.  Fig. (3)).

\begin{figure}[t]
 \centering
  \includegraphics[width=8cm]{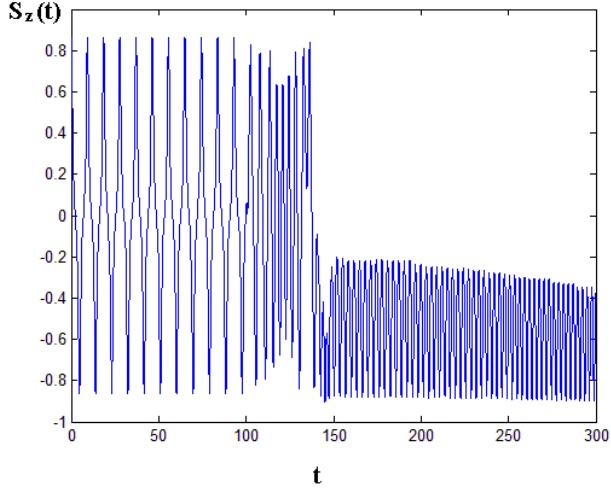}
  \caption{Motion near the separatrix ($k=1$, $\Sigma=\Sigma_{S}=1$), $D=90GHz$, $H=-4.5\cdot10^{3}GHz$, $\varepsilon =0.3$, $\lambda=4$, $\omega_{0}=5$. The
  variable field is applied during the finite time interval between  $\tau_{1}=100$ and $\tau_{2}=150$.
  Before applying  the variable filed, the  motion is regular and is of an oscillatory nature. The variable field
 produces  a transition into the rotary regime and then is switched off. During the transition the motion is chaotic.} \label{Fig:3}
\end{figure}

 \section{Dynamics far from the separatrix: The mean Hamiltonian method}
To conclude our study, finally we consider dynamics far from the
separatrix. The key point is the fact that stochasticity emerges
in the small phase-space domain located near the separatrix. Far
from the separatrix the dynamics is
 regular, even in the presence of small perturbations. In this regime,
 if the frequency of the variable field is high, analytical solutions are found with the
 help of the
 mean Hamiltonian method.
    The basic idea of the mean Hamiltonian method is the following:
for a system having  different time scales, one averages over the
fast variables and obtains thus an explicit expression for the
time independent averaged Hamiltonian \cite{16}. In our case, the
following condition should then
 hold: \be g\mu_{B}H_{0}<D<\omega_{0} \big(\frac{2DS}{\lambda}\big),~~~~~~~
\varepsilon=H_{1}/H_{0}<1.\ee This condition implies that the
amplitude of the magnetic fields  should be small and the frequency should be
 high. Provided  those
conditions  hold it is possible to average the dynamic over the
 fast frequency $\omega_{0}$. The averaged Hamiltonian is
determined by the following expression: \be
H_{av}=\bar{H}+\frac{1}{2}\overline{\{\langle \delta
H\rangle,H\}}+\frac{1}{3}\overline{\{\langle\delta
H\rangle,\{\langle\delta
H\rangle,H+\frac{1}{2}\bar{H}\}\}}+\ldots\ee where $\{A,B\}$ is
the Poisson bracket, $\delta H=H-\bar{H}$, $\langle \delta H
\rangle=\int\delta Hdt$, $\overline{(\ldots)}$ means averaging
over the time. Applying the procedure (25) to the Hamiltonian (1)
and after straightforward  but laborious calculations with the
accuracy up to the second order terms $(1/\omega_{0})^{2}$ we find
\begin{eqnarray}
H_{av}=D
S_{z}^{2}+g\mu_{b}H_{0}\sqrt{1-S_{z}^{2}}\cos\varphi+\frac{1}{2}\bigg[\frac{(g\mu_{B}H_{1})^{2}}{\omega_{0}^2}\times
\nonumber \\
\times
\bigg(-S_{z}^{2}(\cos(\varphi)+\sin(\varphi))^{2}+(1-2S_{z}^{2})(\cos(\varphi)-\sin(\varphi))^{2}\bigg)\bigg].
\end{eqnarray}
The Hamiltonian (26) allows  for further simplification:
Considering
 that the variable $\varphi$ is fast in
comparison with $S_{z}^{2}$, rotating wave approximation can be
used. The Hamiltonian obtained in this way  is completely
identical to (4). This means, that the solutions (9) are still
valid. The difference is that, the constant $\lambda$ has a
different form and depends on the parameters of the variable field

\be
\lambda=\bigg(1-\bigg(\frac{H_{1}g\mu_{B}}{2\omega_{0}}\bigg)^{2}\bigg)\frac{2DS}{g\mu_{b}H_{0}}.
\ee

 By comparing the analytical solutions with the results of the numerical
integration of the system of equations (3) far from the separatrix
we verify the validity of our approximations.

\begin{figure}[t]
 \centering
  \includegraphics[width=16cm]{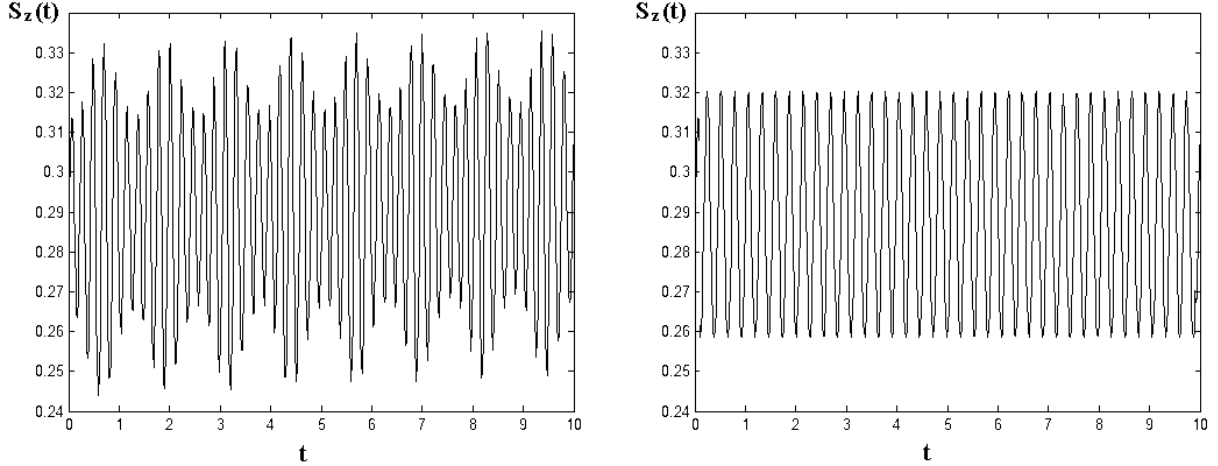}
  \caption{The dynamics far from the separatrix ($k=1.6,~~\Sigma=4\Sigma_{S}$) is  regular;   $D=90GHz$, $H=-0.57\cdot10^{3}GHz$,
    $\varepsilon =0.3$, $\lambda=100$, $\omega_{0}=10$. The
  orientation of the magnetization oscillates with time, however without a change of sign. Dynamically induced switching
  is not possible far from the separatrix. The left plot corresponds to the numerical solution of eq.(3). The right side corresponds to the solution
  (9) $S_{z}(t)=\textrm{bdn}[b\lambda/k(t-\alpha),1/k]$, with re-scaled $\lambda$ constant (27). The solutions are in a good agrement with each other.
  The only difference is the absence of amplitude modulation in the analytical approximation.} \label{Fig:4}
\end{figure}

 Fig.(4) is  for the parameters of the perturbations that are
 analogous to  Fig.(\ref{Fig:2}). However, unlike  Fig.(\ref{Fig:2}),
 where the system is  near the separatrix $k\approx1$, in the case of
 Fig.(4) $k=1.6$ which means that the system is far from the
 separatrix. That is why the dynamics of magnetization is periodic in
 time. The difference, between Fig.4. and the analytical solution
 (9) is that the amplitude of the oscillations is modulated in time. This
 observation can be explained  with the aid of the  average
 Hamiltonian. The point is that the solutions (9) do not
 account for the existence of multiple angles in the average Hamiltonian
 that were ignored by us. They may lead to the appearance of breathing
 and amplitude modulations.

 \section{Conclusions}
We have considered the spin dynamics of a molecular magnet, when
the number of the involved levels  is large. The dynamics  of MM
driven by variable filed has been studied before \cite{12}.
However, in contrast to
 \cite{12}, the applied fields in our case are quite strong,
  i.e. we are in the strongly nonlinear, non-perturbative regime.
The underlying dynamics is then treated semi-classically. We
showed that  the phase space of the system contains two domains
separated by a separatrix line. The solutions far from the
separatrix correspond to the rotating and the oscillatory regime,
 while the separatrix solution is non-oscillating and is of a
soliton type. The existence of  the separatrix is important as
 the states in  the domain near to it are extremely sensitive
to  small perturbations. Therefore, if a variable field is applied,
instead of a soliton type solutions, the spin dynamics turns
 chaotic and unpredictable. The control
parameter is the initial energy of the system. By a proper choice
of it each type of the dynamic can be realized. The structure of
the system's phase space is directly related to the possible
mechanisms of the magnetization reversal. Namely, if the energy is
equal to $H_{S}=\frac{8}{9} E_{B}^{'}$ of the re-scaled anisotropy
barrier $E_{B}^{'}=DS^{2}\big(1-\frac{1}{\lambda}\big)^{2}$
\cite{11} (the separatrix condition) an external variable field
leads to a chaotic change of the magnetization orientation. The
switching process is random and with the equal probability 1/2,
the system may appear in the new state as well as stay in the old
one. The information about initial state is lost. This result is
different from the case of weak applied fields \cite{12}, where
the dynamics shows a long-term memory of the initial state.

\textbf{Acknowledgment:} The  project is financially supported by
 the Georgian National Foundation
(grants: GNSF/STO 7/4-197, GNSF/STO 7/4-179). The financial
support by the Deutsche Forschungsgemeinschaft (DFG) through SFB
672 and though  SPP 1285  is gratefully acknowledged.

\end{document}